\documentclass[twocolumn,pra,nobalancelastpage,aps,raggedbottom, superscriptaddress]{revtex4-2}
\usepackage{times}

\usepackage{amsmath,amsfonts,amssymb,graphicx,hyperref,epstopdf,xcolor,tikz,scalerel,natbib,array}
\usepackage{mathptmx,textcomp,braket}
\usepackage[T1]{fontenc}
\usepackage[utf8]{inputenc}
\usepackage{multirow}
\usepackage{array}
\usepackage{tabularx}
\usepackage{makecell}

\definecolor{lime}{HTML}{A6CE39}
\DeclareRobustCommand{\orcidicon}
{
	\begin{tikzpicture} 
	\draw[lime, fill=lime] (0,0) circle [radius=0.15] node[white] {{\fontfamily{qag}\selectfont \tiny ID}};
	\draw[white, fill=white] (-0.0625,0.095) 	circle [radius=0.007];
	\end{tikzpicture}
	\hspace{-2.2mm}
}
\newcommand\orcidID[1]{\href{https://orcid.org/#1}{\orcidicon}}

\newcommand{\be}{\begin {equation}}
\newcommand{\ee}{\end {equation}}
\newcommand{\beqa}{\begin {eqnarray}}
\newcommand{\eeqa}{\end {eqnarray}}
\newcommand{\mb}{\mathbf}
\newcommand{\Sch}{Schr\"odinger }

\hypersetup{colorlinks,citecolor=blue,filecolor=black,linkcolor=blue,urlcolor=blue}

\begin{document}

\title{Systematic analysis of an attosecond pulse generation by a sub-cycle laser field}

\author{Rambabu Rajpoot\orcidID{0000-0002-2196-6133}}
\email[E-mail: ]{rambabu.rajpoot@riken.jp}
\affiliation{RIKEN Center for Advanced Photonics, RIKEN, 2-1 Hirosawa, Wako, Saitama 351-0198, Japan}
\affiliation{RIKEN Cluster for Pioneering Research, RIKEN, 2-1 Hirosawa, Wako, Saitama 351-0198, Japan}

\author{Eiji J. Takahashi\orcidID{0000-0002-6287-2969}}
\email[Corresponding author: ]{ejtak@riken.jp}
\affiliation{RIKEN Center for Advanced Photonics, RIKEN, 2-1 Hirosawa, Wako, Saitama 351-0198, Japan}
\affiliation{RIKEN Cluster for Pioneering Research, RIKEN, 2-1 Hirosawa, Wako, Saitama 351-0198, Japan}

\date{\today}

\begin{abstract}
We investigated the influence of sub-cycle driving fields on high-order harmonic generation (HHG), with a focus on intrinsic chirp, carrier-envelope phase (CEP), and number of laser cycles. Our findings reveals that the center frequency of a laser pulse scales as $\tau^{-5/4}$ with pulse duration $\tau$, and that attochirp exhibits a similar dependence on pulse duration. Additionally, we identified CEP-specific trends in harmonic yield: it increases as $\tau^{5/4}$ for $\phi_0=0^\circ$ and decreases as $\tau^{-4.1}$ for $\phi_0= -90^\circ$. Although sub-cycle pulses can generate intense isolated attosecond pulses (IAPs), they also tend to produce higher attochirp and reduced cutoff energies. However, effective compensation for attochirp can mitigate these drawbacks, thereby increasing the capability of sub-cycle pulses to generate short-duration, high-intensity IAPs. These results offer valuable insights into HHG using sub-cycle pulses and have important implications for the advancement of ultrafast light sources and the understanding of ultrafast phenomena at the attosecond timescale. 
\end{abstract}

\maketitle

\section{Introduction}

High-order harmonic generation (HHG) has become a vital mechanism for producing extreme-ultraviolet (XUV) and soft X-ray radiation that possesses high spatial and temporal coherence \cite{Liu2019SpecLett,Mairesse2003_Science}. The harmonic generation process is effectively captured by the \textit{three-step model} \cite{PhysRevLett.70.1599,Corkum1993_PRL}, which includes electron ionization, accelerated in the laser field, and recombination with the parent ion, resulting in the emission of higher harmonics of the driving laser. The fundamental aspects of HHG have been analytically investigated through quantum electron dynamics \cite{Frolov2009_scale, Katalin2010_Quantum, Astiaso2016, PhysRevLett.95.223901}. These higher harmonics play an essential role in generating attosecond pulses, which are applicable across a broad range of fields \cite{Krausz2009_RMP, Hentschel2001_Nature, Corkum2007_NatPhy, Chini2014_nat, Heuser2016_PRA, Ayuso2018_JPhysB, Baykusheva2016_PRL, Reich2016_PRL}. 

Recent developments in both theoretical and experimental fields have made the generation of sub-cycle laser pulses possible \cite{Xu2024_NatPhoton, Nishimiya2024_OptLett, Ding-2015, Takahashi-2013, Voronin2013_OptLett, Shverdin2005_PRL, Emma-2004, Brabec2000_RevModPhys}, resulting in increased interest in exploring quantum dynamics with such ultra-short pulses \cite{Rossi2020, Chu_2016, Liang2017, Goulielmakis2008_Science}.For laser pulses longer than a single cycle, the field profiles are effectively represented by carrier-envelope (CE) expressions. However, in the sub-cycle regime, these expressions become insufficient, leading to unrealistic field profiles \cite{Brabec1997_PRL}. Specifically, the field profile described by the CE expression acquires a DC component that can not be linked to a propagating electromagnetic pulse, and the spectrum of the pulse changes with its carrier-envelope phase (CEP). 

To address this issue, an analytical expression for laser pulses of arbitrary duration has been derived from the oscillating dipole model \cite{LinSubcycle-2006}. The resulting field profiles are exact solutions to Maxwell's equations and are produced using the complex-source point method \cite{PhysRevE.67.016503,Heyman-89}. This analytical expression includes a temporal phase shift, similar to the spatial Gouy phase observed in focused beams, leading to the prediction of an intrinsic chirp that induces a blueshift in the center frequency of sub-cycle pulses \cite{LinSubcycle-2006}. This intrinsic chirp has been experimentally confirmed in sub-cycle terahertz pulses through time-frequency analysis \cite{Lin2010_PRA} and carries considerable implications for strong-field physics. For example, it affects the energy gain of relativistic electrons \cite{LinSubcycle-2006} and contributes to the self-steepening of sub-cycle pulses in nonlinear media \cite{Cai2016_OptCommun}.

The waveform of the laser electric field is crucial in determining the characteristics of HHG. The time-dependent frequency of emitted harmonics is closely associated with the chirp of the driving laser, which, subsequently influences the chirp of the resulting attosecond pulses \cite{Kazamias2004_PRA, Murakami2005_PRA, Murakami2005_PhysRevA}. Therefore, accounting for the intrinsic chirp of sub-cycle pulses is essential when generating high harmonics with such short-duration pulses. 

Previous studies have used sub-cycle pulses to examine the effects of pulse envelopes on harmonic cutoffs \cite{Zheng_2011} and the scaling laws of harmonic yield and cutoff energy \cite{Holkundkar2023_PLA}. In this work, we investigate HHG driven by sub-cycle laser pulses interacting with a helium atom, specifically focusing on the effect of the intrinsic chirp of the sub-cycle pulse on the chirp of the generated attocecond pulses and the influence of the sub-cycle pulse CEP on the harmonic yield.

The paper is structured as follows. Section \ref{sec2} outlines the numerical methods, followed by the results and discussion provided in Section \ref{sec3}. Section \ref{sec4} offers concluding remarks and future directions. Unless otherwise specified, atomic units (a.u.) are used throughout.
  
\section{Numerical Methods}
\label{sec2} 

The study is performed by numerically solving the time-dependent \Sch equation (TDSE) under a single-active-electron approximation using the time-dependent generalized pseudospectral method \cite{TONG1997119}. The TDSE in the length gauge is formulated as follows:
\be
	i \frac{\partial}{\partial t} \psi(\mb{r},t) = \big[- \frac{1}{2} \nabla^2 + V(r) + \mb{r}.\mb{E}(t) \big] \psi(\mb{r},t) ,
 \ee
where $\mb{E}(t)$ denotes the temporal profile of the linearly polarized laser pulse in the dipole approximation. Once the TDSE is solved, the harmonic power spectrum is derived by applying the Fourier transform to the dipole acceleration $\mb{a}(t)$ as follows,

\be S(\omega) = \frac{1}{\omega^4} \Big|\frac{1}{\sqrt{2\pi}} \int \mb{a}(t) e^{-i\omega t} dt \Big|^2 = \frac{|a(\omega)|^2}{\omega^4}. \ee

The integrated harmonic yield is calculated as \cite{Ishikawa2009_scale,Schiessl2007_scale}, 
\be Y = \frac{1}{\tau} \int_{\epsilon_i}^{\epsilon_f} |a(\omega)|^2 d\omega, \label{yield}\ee
where $\tau$ is the full-width at half-maximum (FWHM) pulse duration, and $[\epsilon_i:\epsilon_f]$ is the chosen harmonic bandwidth in the plateau region. Additionally, the field profile of the attosecond pulse, $\mathcal{E}_{asp}(t)$ can be constructed by filtering the desired frequency range with an appropriate window function $w(\omega)$, as given in \cite{Peng-2020},
\be \mathcal{E}_{asp}(t) = \frac{1}{\sqrt{2\pi}} \int a(\omega) w(\omega) e^{i\omega t} d\omega,
\label{E_asp} \ee 
here, $w(\omega) = \Theta(\omega-\omega_1) \Theta(\omega_2-\omega)$, where $\omega_1 \leq \omega \leq \omega_2$ is the frequency range to be filtered, and $\Theta(x)$ is the standard step function. The intensity of the attosecond pulse is then expressed as $I(t) \sim |\mathcal{E}_{asp}(t)|^2$. 

To describe the laser pulse with an arbitrary envelope and pulse length, we use the analytical expression of a sub-cycle pulsed beam (SCPB) as presented in Ref. \cite{LinSubcycle-2006}. SCPBs are exact solutions to Maxwell's equations and derived from the oscillating dipole model through the complex-source point method \cite{PhysRevE.67.016503,Heyman-89}. The expression for a linearly polarized sub-cycle pulse in the plane-wave limit is given as,
\be
\mb{E}(t) = \text{Re}\bigg\{ \frac{A_d(t')}{|A_d(0)|} f(t')\ E_0\exp(i\omega_0t' + i\phi_0) \bigg\} \hat{\mb{e}}_z,
\label{laser}
\ee
where $t' = t - t_0 - z/c$ is the retarded time (with $z = 0$ and $t_0$ denoting the pulse center), $E_0$ is the peak field amplitude, $\omega_0$ is the carrier frequency, and $\phi_0$ represents the CEP. The complex time-dependent function $A_d(t')$ is defined as:
\be
A_d(t') = 1 + [i\omega_0 f(t')]^{-1} \dot{f}(t'),
\ee 
where $\dot{f}(t')$ is the first-order derivative of the envelope function $f(t')$ with respect to the retarded time $t'$. The pulse envelope $f(t')$ can be chosen arbitrarily. However, to obtain a laser pulse with intensity that approaches zero at the ends of the pulse waveform, the following conditions must be satisfied \cite{Zheng_2011}:
\be
 \lim_{t' \rightarrow\pm \infty} f(t') = 0,\quad  \lim_{t' \rightarrow\pm \infty} \dot{f}(t') = 0, \quad \lim_{t' \rightarrow\pm \infty} \ddot{f}(t') = 0 
\ee 
In this work, we relied on the analytical Gaussian envelope to model the sub-cycle pulses. The envelope function is defined as \cite{Cai2016_OptCommun}:
\be
f(t') = f_0\ e^{-(t'/T)^2} \Big(1+ \frac{2i}{\sqrt{\pi}T} \int_0^{t'} e^{(t''/T)^2} dt'' \Big),
\ee
where, $f_0 = \big(1+ \frac{2i}{\sqrt{\pi}T} \int_0^{t'} e^{(t''/T)^2} dt'' \big)^{-\frac{1}{2}}$ is the normalization factor used to prevent the broadening of the envelope, and $T$ is the pulse width parameter associated with the FWHM $\tau$ ($= 2\sqrt{\text{ln}2}T$) of the pulse envelope function $f(t')$. 
 
In the simulation, we consider a linearly polarized laser with center wavelength $\lambda_0 = 1600$ nm and peak intensity $I_0 = 8\times 10^{14}$ W$/$cm$^{2}$ (unless otherwise noted). The total simulation time is taken to be $4\tau$, where $\tau$ is defined as the FWHM of the envelope function $f(t')$. The atomic Coulomb potential $V(r)$ for helium atom under single-active-electron approximation is modeled using an empirical expression \cite{Tong_2005}, wherein the coefficients of $V(r)$ are obtained by the self-interaction free density functional theory. The ground state (initial state) energy of the helium is determined to be $\sim -0.9038$ a.u. A radial simulation domain of 250 a.u. is adopted with the last 20 a.u. utilized as a masking region to absorb the outgoing wave-function \cite{TONG1997119}. Additionally, a maximum angular momentum of $\sim 220$ is considered, and a simulation time step of $\sim 0.05$ a.u. is used. Convergence is tested with respect to the spatial grid and the time step. Our simulation uses the  \textit{Armadillo} library for linear algebra purposes \cite{Sanderson2016}.

In the following sections, we examine the harmonic generation by sub-cycle pulses, with a focus on the effects of intrinsic chirp and CEP on the resulting harmonics.

\section{Results and Discussion}
\label{sec3}
\begin{figure}[t]
\centering\includegraphics[width=1\columnwidth]{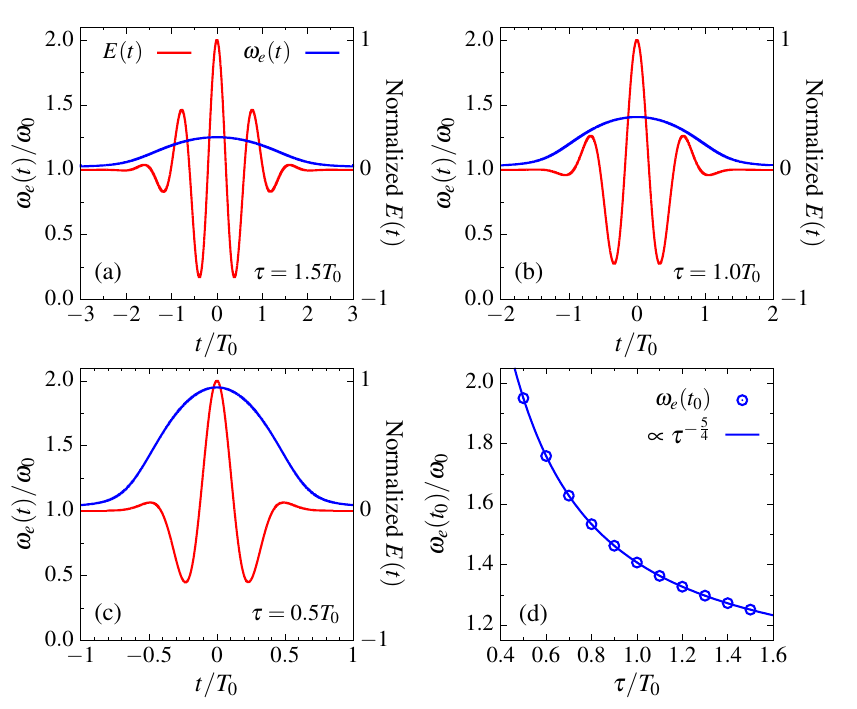}
\caption{Temporal profiles of the instantaneous frequency and electric field of the sub-cycle pulse for different FWHM durations: (a) $\tau = 1.5T_0$, (b) $\tau = 1.0T_0$, and (c) $\tau = 0.5T_0$. (d) Relative blueshift of the sub-cycle pulse frequency at the pulse center ($t_0=0$) for different FWHM durations. The frequency blueshift scales as $\propto \tau^{-5/4}$ with the pulse duration $\tau$. The points indicate computed results, while the solid line represents the scaling. Here, $T_0\ (= 2\pi/\omega_0)$ denotes one laser cycle. }
\label{fig1}
\end{figure}

We begin by analyzing the dependence of laser's instantaneous frequency on the pulse duration. The total phase of the pulse defined in Eq. (\ref{laser}), is given as:
\be \phi(t') = \omega_0 t' + \phi_0 + \text{arg}[f(t')] + \text{arg}[A_d(t')], 
\label{total_phase}
\ee 

The corresponding instantaneous frequency of the pulse can be obtained by taking the derivative of the total phase as $\omega_e(t') = d\phi(t')/dt'$. This time-dependence of laser frequency results in a chirped pulse. Because this chirp originates from the finite nature of the pulse rather than from material dispersion, it is referred to as intrinsic chirp \citep{Lin2010_PRA}. Panels (a)-(c) of Fig. \ref{fig1} display the instantaneous frequency of a sub-cycle plane wave defined in Eq. (\ref{laser}), alongside the temporal waveform of a Gaussian pulse for various pulse durations. It is evident that the time-dependent frequency is symmetric in the pulse and is blueshifted with respect to the carrier frequency $\omega_0$ throughout the entire pulse. Notably, the pulse exhibits nonlinear chirping, with the center frequency considerably surpassing the carrier frequency $\omega_0$, reaching approximately $2\omega_0$ for a $\tau-0.5T_0$ duration pulse, as illustrated in Fig. \ref{fig1}(c). Figure \ref{fig1}(d) shows the blueshift of the center frequency $\omega_e(t_0)$ owing to intrinsic chirp as a function of FWHM duration $\tau$. It is observed that the center frequency scales as $\propto \tau^{-5/4}$ with varying pulse duration $\tau$. The intrinsic chirp becomes increasingly pronounced as the pulse duration transitions from a few-cycle to the sub-cycle regime. The presence of intrinsic chirp manifests in shrinking the waveform of sub-cycle pulses. Considering the critical role of the laser electric field waveform in high harmonic generation, accounting for the intrinsic chirp of sub-cycle pulses is essential when generating high harmonics with such short-duration pulses. The time-dependent frequency of the emitted harmonics is closely linked to the chirp of the driving laser, which subsequently affects the chirp of the resulting attosecond pulses. Therefore, it is intriguing to explore how the attochirp changes as the duration of the sub-cycle driving pulse changes.

\subsection{Analysis of the characteristics of high-order harmonic generation}
\label{subsec1}

\begin{figure}[t]
\centering\includegraphics[width=1\columnwidth]{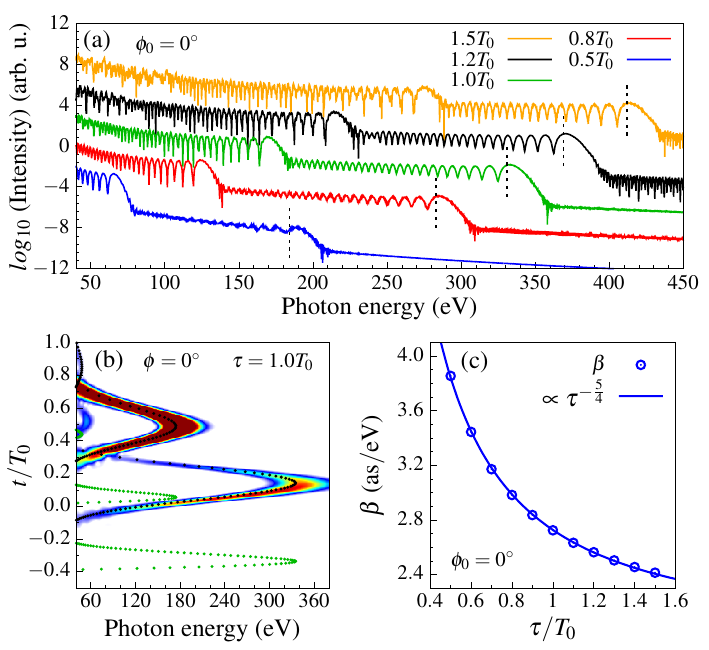}
\caption{ (a) High harmonic spectra generated for different driving pulse durations. The vertical dashed-lines indicating the cutoff energies estimated using the modified expression. For clarity, the HHG spectra for $\tau = 0.8T_0,\ 1.0T_0,\ 1.2T_0,$ and $1.5T_0$ are shifted along the y-axis by factors $10^{3},\ 10^{6},\ 10^{9},$ and $10^{12}$, respectively. (b) The time-frequency profile of the dipole acceleration for a pulse duration of $\tau = 1.0T_0$ is presented. The corresponding classical re-collision energies for ionization (green dots) and recombination (black dots) times are also shown. (c) Scaling of the attochirp $\beta$ as a function of pulse duration. The attochirp follows a scaling of $\propto \tau^{-5/4}$. The points represent computed results and solid line represents scaling. } 
\label{fig2}
\end{figure}

To explore the impact of sub-cycle pulse intrinsic chirp on the temporal structure of resulting attosecond pulses, we first examine the corresponding high harmonic spectra generated by these short pulses. Figure \ref{fig2}(a) displays the HHG spectra of helium for various pulse durations, with the CEP $\phi_0$ fixed at $0^\circ$ in all cases. As the pulse duration decreases from $1.5T_0$ to $0.5T_0$, the harmonic cutoff energy is reduced by nearly half.

This reduction in cutoff energy can be explained as follows. The photon energy at the cutoff is given by the relation $E_{\text{c}} = I_p + 3.17U_p$, where $I_p$ is the ionization potential of the atom, and $U_p= E^2/(2\omega)^2$ is the ponderomotive quiver energy of the electron in the laser field, with amplitude $E$ and frequency $\omega$. For sub-cycle pulses, the strength of successive electric field extrema varies significantly. The electron quiver energy is predominantly determined by the strength of the electric field extremum $E_{\text{ext}}$ in the returning region \cite{Haworth2007_NatPhys, Chipperfield2009_PRL}. Additionally, the center frequency $\omega_e (t_0)$ of the sub-cycle pulse more accurately characterizes the interaction between these ultra-short pulses and matter \cite{Neyra2021_PRA}. 
Therefore, a more accurate estimation of the harmonic cutoff energy can be achieved by modifying the cutoff energy expression as follows, 
\be
 E_{\text{sc}}\ [\text{eV}] = I_p + \frac{9.33 I_{\text{ext}} \lambda_0^2}{ (1+\alpha\tau_0^{-5/4})^{2}}, 
\label{new_cutoff} 
\ee
where $\lambda_0$ is the carrier wavelength in $\mu$m, $\tau_0$ represents the pulse duration in units of the laser cycle $T_0$, and $\alpha=0.4$ is a fitting constant determined from the center frequency scaling depicted in Fig. \ref{fig1}(d). The parameter $I_{\text{ext}}\ (\sim |E_{\text{ext}}|^2)$ denotes the laser intensity at the field extremum $E_{\text{ext}}$, given in W$/$cm$^2$. For a CEP of $\phi_0=0^\circ$ (cosine-like pulses), $I_{\text{ext}}$ corresponds to the peak intensity $I_0$ of the driving pulse. For instance, in the $0.5T_0$ pulse duration HHG spectra shown in Fig. \ref{fig2}(a) (blue curve), the parameters are as follows: $I_p = 24.6$ eV (for a He atom), $\lambda_0 = 1.6 \mu$m, $\tau_0 = 0.5$, and $I_{\text{ext}} = 8\times 10^{14}$ W$/$cm$^2\ (\sim I_0)$ for $\phi_0 = 0^\circ$. The estimated harmonic cutoff energy is $E_{\text{sc}} \sim 184$ eV.
In Fig. \ref{fig2}(a), the harmonic cutoff energies calculated using the modified expression (Eq. \ref{new_cutoff}) are marked by vertical dashed lines. As the pulse duration $\tau$ decreases, the center frequency of the sub-cycle pulse increases approximately as $\tau^{-5/4}$, as illustrated in Fig. \ref{fig1}(d). This results in a corresponding reduction in the ponderomotive energy, which explains the observed decrease in harmonic cutoff energy. Furthermore, the contribution of the continuum harmonic region (i.e., the range between the first and second cutoff energies) to the total HHG spectrum grows as the pulse duration decreases. Specifically, it increases from about $35\%$ of the total HHG spectrum for a duration of $\tau=1.5T_0$ to approximately $65\%$ for the $\tau=0.5T_0$ case.The broader continuum harmonic region is essential for generating short-duration isolated attosecond pulse (IAP).

The attochirp is best characterized by the slope of the curve representing emission times as a function of harmonic energy. Figure \ref{fig2}(b) illustrates the instants of harmonic emission across the spectrum for a $\tau = 1.0T_0$ duration driving pulse [described in Fig. \ref{fig1}(b)]. The time-frequency profile is derived from the Gabor transformation of the dipole acceleration \cite{Gabor1946, Arts2022_NatComputSci}. Additionally, the classical ionization (green) and recombination (black) energies corresponding to electron trajectories driven by the single-cycle pulse are also shown. As is typically observed in HHG calculations, the classical energies closely follow the temporal evolution of harmonic emission obtained from quantum mechanical TDSE solutions. The dependence of the attochirp $\beta$ on the sub-cycle driver length can thus be understood using simple Newtonian calculations. Figure \ref{fig2}(c) presents the variation in the attochirp as a function of the pulse duration $\tau$. The attochirp is calculated for the short trajectory with the highest returning energy using classical trajectory analysis. It is observed that the attochirp scales as $\propto \tau^{-5/4}$ with the pulse duration. 

The rationale for this particular scaling of attochirp with pulse duration can be given as follows. Attochirp $\beta$ is defined as the ratio of emission time to harmonic energy. For a sub-cycle laser pulse, the emission time is proportional to the laser period $T_e\ (=2\pi/\omega_e)$, while the harmonic energy is proportional to $U_p$, which decreases as $\omega_e^{-2}$. Therefore, the attochirp $\beta \propto T_e/U_p$ is proportional to the blueshifted center frequency $\omega_e$. As shown in Fig.\ref{fig1}(d), the center frequency scales as $\propto \tau^{-5/4}$, leading to the attochirp also following the same scaling of $\beta \propto \tau^{-5/4}$ with the pulse duration $\tau$. 

\begin{figure}[t]
\centering\includegraphics[width=1\columnwidth]{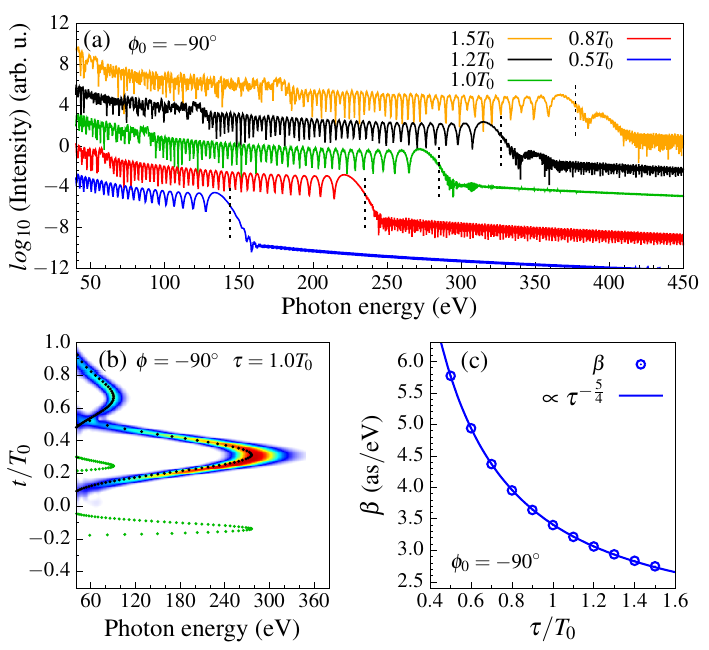}
\caption{ Similar to Fig. \ref{fig2}, but with CEP $\phi_0 = -90^\circ$. (a) HHG spectra generated for different driving pulse durations. The HHG spectra for $\tau = 0.8T_0,\ 1.0T_0,\ 1.2T_0,$ and $1.5T_0$ are shifted along the y-axis by factors $10^{3},\ 10^{6},\ 10^{9},$ and $10^{12}$, respectively. (b) Time-frequency profile of the dipole acceleration for $\tau = 1.0T_0$. (c) Scaling of the attochirp $\beta$ as a function of pulse duration, following the same $\propto \tau^{-5/4}$ scaling.} 
\label{fig3}
\end{figure}

To verify the universality of attochirp scaling, we calculated the HHG spectra and the corresponding attochirp for various pulse durations with the CEP $\phi_0 = -90^\circ$, as shown in Fig.\ref{fig3}. The driving field, having a sine-like waveform ($\phi_0 = -90^\circ$), produces higher yield and lower cutoff energy of secondary plateau harmonics that originate from electron excursions during the main cycle of the driving pulse, as evident from the time-frequency profile in Fig. \ref{fig3}(b). Notably, the HHG spectrum shows a dominant contribution from the short trajectory harmonics. The calculated attochirp $\beta$ for these short trajectory harmonics, along with its scaling with pulse duration $\tau$, is presented in Fig. \ref{fig3}(c). The attochirp $\beta$ exhibits the same scaling of $\propto \tau^{-5/4}$ with pulse duration. Considering the significance of CEP in ultrashort driving fields \cite{Bohan1998_PRL, Nisoli2003_PRL, Haworth2007_NatPhys}, this consistency of attochirp scaling, despite variations in CEP $\phi_0$, confirms the universality of attochirp scaling with pulse duration. 

Moreover, a comparison of the scalings in panel (c) of Figs. \ref{fig2} and \ref{fig3} reveals that while the overall scaling behavior remains consistent, the amplitude of the attochirp $\beta$ is greater for $\phi_0 = -90^\circ$ than the $\phi_0 = 0^\circ$ case. This increased amplitude of attochirp is attributed to the reduction in harmonic cutoffs observed in the $\phi_0 = -90^\circ$ HHG spectra, compared to those at $\phi_0 = 0^\circ$, generated by the laser pulse of the equal duration.

\begin{figure}[t]
\centering\includegraphics[width=1\columnwidth]{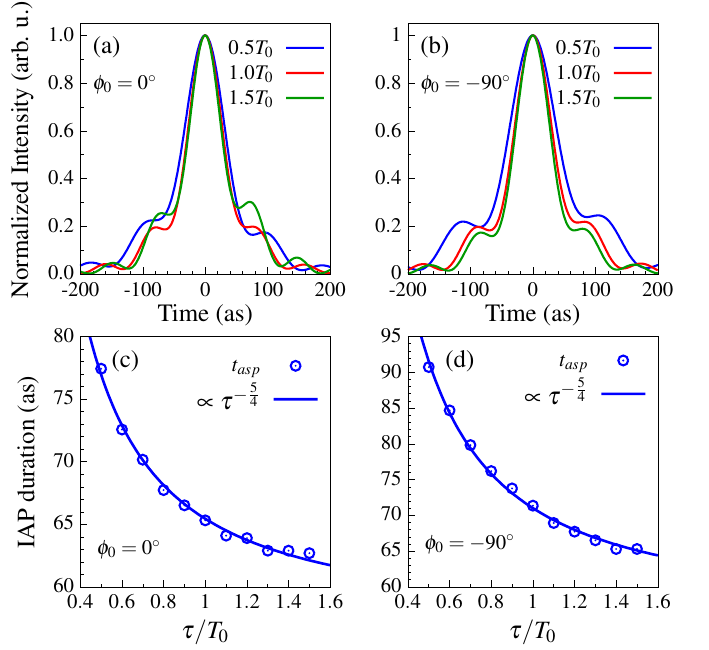}
\caption{ Temporal profiles of isolated attosecond pulses (IAPs) generated for CEP values $\phi_0$: (a) $0^\circ$, (b) $-90^\circ$. The IAPs corresponding to different pulse durations are normalized to their respective peak intensities and shifted in time for comparison. The time on the x-axis is given in attoseconds (as). Panels (c) and (d) show the scaling of IAP duration $t_{asp}$ as a function of pulse duration for $\phi_0$: (c) $0^\circ$, (d) $-90^\circ$. The IAP duration follows a consistent scaling of $\propto \tau^{-5/4}$. } 
\label{fig4}
\end{figure}

To visualize the effect of intrinsic chirp induced attochirp, we synthesized attosecond pulses by superposing short trajectory harmonics generated for various pulse durations, while keeping the CEP $\phi_0$ fixed at $0^\circ$ and $-90^\circ$. The harmonic window [refer to Eq. (\ref{E_asp})] was carefully selected to produce the shortest isolated attosecond pulse (IAP) (without any phase compensation) after scanning through the entire plateau harmonics. The intensity profiles of the synthesized IAPs for laser durations $\tau = 0.5T_0$, $1.0T_0$, and $1.5T_0$, with CEP $\phi_0$ fixed at $0^\circ$ and $-90^\circ$, are presented in Figs. \ref{fig4}(a) and \ref{fig4}(b), respectively. The selected harmonic windows corresponding to these three pulse durations are detailed in Table \ref{tab_IAPwindow}. The effect of attochirp variation with pulse duration is reflected in the width of the harmonic window that supports the shortest IAP for both the CEP cases. Specifically, the harmonics bandwidth for CEP $\phi_0=0^\circ (-90^\circ)$ decreases from $86 (79)$ harmonics to $69(57)$ harmonics as the pulse duration changes from $1.5T_0$ to $0.5T_0$. The variation of IAP duration $t_{asp}$ with driver length $\tau$ for the two CEP values is illustrated in Figs. \ref{fig4}(c) and \ref{fig4}(d). As expected, the duration $t_{asp}$ consistently follows the scaling of $\propto \tau^{-5/4}$ with pulse duration, irrespective of the CEP value. Notably, the IAPs generated with $\phi_0 = -90^\circ$ exhibit longer durations than those generated with $\phi_0 = 0^\circ$. This increase in IAP duration can be attributed to the higher attochirp values associated with $\phi_0 = -90^\circ$, as discussed earlier. Furthermore, for $\phi_0 = -90^\circ$, the IAP duration increases by approximately $40\%$ as the pulse duration decreases from $1.5T_0$ to $0.5T_0$, highlighting the effect of intrinsic chirp associated with the sub-cycle driving fields. At this stage, it is intriguing to consider how the yield of generated HHG spectra changes as the duration of the driving pulse varies from a few cycle to the sub-cycle regime.

\begin{table}[b]
\setlength{\tabcolsep}{2pt}
\renewcommand{\arraystretch}{1.3}
\centering
\begin{tabular}{p{1.2cm} p{2.3cm} p{2.3cm} p{1.7cm} }
\hline\hline
\multirow{2}{*}{$\tau$ ($T_0$)} & $\omega_1/\omega_0$ & $\omega_2/\omega_0$ & $\Delta \omega/\omega_0$ \\ \cline{2-4} 
     & \hspace{-0.3cm}$\phi_0 = 0^\circ$ ($-90^\circ$) & \hspace{-0.3cm}$\phi_0 = 0^\circ$ ($-90^\circ$) & \hspace{-0.3cm}$\phi_0 = 0^\circ$ ($-90^\circ$)  \\ \hline
0.5 & 126 (102) & 195 (159) & 69 (57)  \\ 
1.0 & 240 (176) & 321 (249) & 81 (73) \\ 
1.5 & 397 (269) & 483 (348) & 86 (79) \\ 
\hline\hline
\end{tabular}
\caption{Calculated harmonic-order window corresponding to shortest IAP for different pulse durations with CEP values $\phi_0 = 0^\circ$ and $-90^\circ$.}
\label{tab_IAPwindow}
\end{table}

\subsection{Scaling of harmonic yield}
\label{subsec2}
\begin{figure}[t]
\centering\includegraphics[width=1\columnwidth]{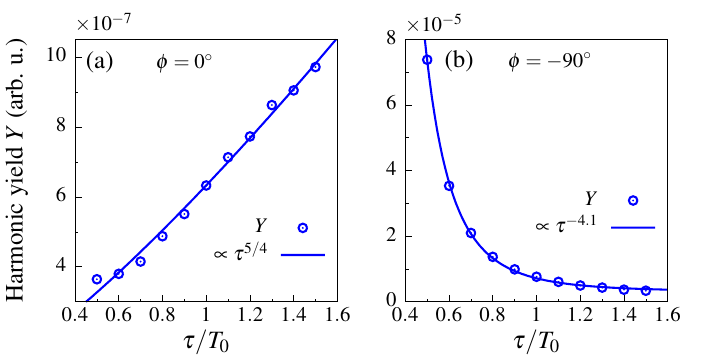}
\caption{Scaling of integrated harmonic yield as a function of laser pulse duration for CEP values $\phi_0$: (a) $0^\circ$, (b) $-90^\circ$. The harmonic yield is computed for an energy window of 60 eV. For $\phi_0=0^\circ$, the yield increases as $\propto \tau^{5/4}$ with pulse duration, whereas for $\phi_0=-90^\circ$, it decreases following a scaling of $\propto \tau^{-4.1}$. The points represent computed results, and the solid lines illustrate the respective scaling trends.} 
\label{fig5}
\end{figure}

So far, we have established that the attochirp $\beta$ is sensitive to the duration of driving laser pulse, exhibiting a scaling of $\beta \propto \tau^{-5/4}$ with pulse duration. Next, we will analyze how the harmonic yield varies with the duration of the driving pulse. In Fig. \ref{fig5}, the integrated harmonic yield $Y$ [refer Eq. \ref{yield}] is presented for CEP values of $0^\circ$ and $-90^\circ$, with pulse durations ranging from $0.5T_0$ to $1.5T_0$. Generally, in harmonic yield calculations, the harmonic photon energies $\epsilon_i$ and $\epsilon_f$ are kept fixed along with the photon energy window $\Delta\epsilon$ ($=\epsilon_f - \epsilon_i$). However, in this case, due to the multi-plateau structure of the HHG spectra [see Fig. \ref{fig2}(a)], it is not feasible to keep $\epsilon_i$ and $\epsilon_f$ constant over the entire pulse duration range from $0.5T_0$ to $1.5T_0$. To mitigate possible fluctuations due to the multi-plateau nature of the spectra, we increased the harmonic energies $\epsilon_i$ and $\epsilon_f$ by $20$ eV for every $0.1T_0$ increase in pulse duration, while maintaining the energy window $\Delta\epsilon$ fixed at $60$ eV. The corresponding harmonic energy values for different pulse durations are summarized in Table \ref{tab1}. Two distinct trends in the scaling of integrated harmonic yield with pulse duration were observed for the different CEP cases. For $\phi_0 = 0^\circ$ [Fig. \ref{fig5}(a)], the harmonic yield increases as $\propto \tau^{5/4}$ with pulse duration. In contrast for $\phi_0 = -90^\circ$ [Fig. \ref{fig5}(b)], the yield decreases, following a scaling of $\propto \tau^{-4.1}$. This underscores the considerable impact of CEP on the harmonic generation process in the context of sub-cycle driving fields.

\begin{table}[b]
\setlength{\tabcolsep}{10pt}
\renewcommand{\arraystretch}{1.3}
\centering
\begin{tabular}{p{1.5cm} m{2.0cm} p{1.7cm} }
\hline\hline
\multirow{2}{*}{$\tau$ ($T_0$)} & $\epsilon_i$ (eV) & $\epsilon_f$ (eV) \\ \cline{2-3} 
     & \hspace{-0.3cm}$\phi_0 = 0^\circ$ ($-90^\circ$) & \hspace{-0.3cm}$\phi_0 = 0^\circ$ ($-90^\circ$)  \\ \hline
0.5 & 110 (50) & 170 (110)  \\ 
0.7 & 150 (90) & 210 (150)  \\ 
0.9 & 190 (130) & 250 (190)  \\ 
1.1 & 230 (170) & 290 (230)  \\ 
1.3 & 270 (210) & 330 (270)  \\ 
1.5 & 310 (250) & 370 (310)  \\ 
\hline\hline
\end{tabular}
\caption{Harmonic photon energy values for various pulse durations with CEP values $\phi_0 = 0^\circ$ and $-90^\circ$.}
\label{tab1}
\end{table}

\begin{figure}[t]
\centering\includegraphics[width=1\columnwidth]{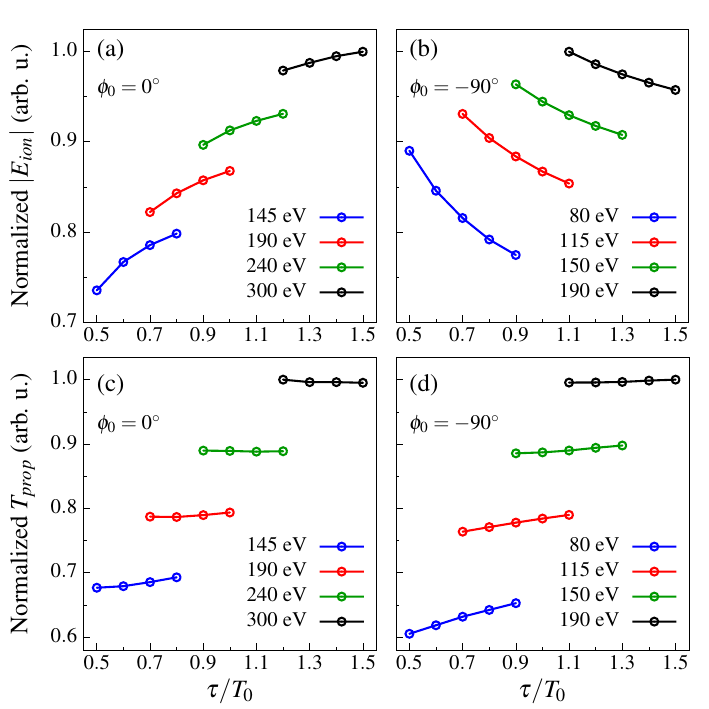}
\caption{Dependence of the laser field strength on sub-cycle pulse duration for different harmonic energies at CEP values of $\phi_0$: (a) $0^\circ$, (b) $-90^\circ$. The classical electron excursion times for the same harmonic energies are shown for $\phi_0$: (c) $0^\circ$, (d) $-90^\circ$. In each panel, all data points are normalized to the maximum value within that panel.}
\label{fig6}
\end{figure}

The harmonic yield for a given target atom is primarily influenced by the strength of the driving field and the spreading of the continuum electron wave packet, which depends on the electron's excursion time in the continuum. To further understand how the number of cycles of the pulse affects harmonic yield, we calculated the classical electron propagation time ($T_{\text{prop}}$) and the electric field strength ($E_{\text{ion}}$) at the moment of ionization using classical trajectory analysis. Given that short trajectory harmonics dominate the HHG spectra, as shown in Figs. \ref{fig2}(b) and \ref{fig3}(b), our analysis focuses on the field strengths $E_{\text{ion}}$ and excursion times $T_{\text{prop}}$ for these harmonics.

First, we consider the case of CEP $\phi_0 = 0^\circ$, shown in Figs. \ref{fig6}(a) and \ref{fig6}(c). Here, the field strength $E_{\text{ion}}$ increases steadily with pulse duration for fixed laser intensity $I_0 = 8\times 10^{14}$ W$/$cm$^2$, while the propagation time $T_{\text{prop}}$ remains nearly constant for a given harmonic energy. This increase in field strength leads to a corresponding increase in harmonic yield, as seen in Fig. \ref{fig5}(a). In contrast, for the $\phi_0 = -90^\circ$ case, shown in Figs. \ref{fig6}(b) and \ref{fig6}(d), the field strength $E_{\text{ion}}$ decreases with pulse duration, while the propagation time $T_{\text{prop}}$ shows minimal variation. This results in a decreasing harmonic yield as the pulse duration increases. Furthermore, for longer pulse durations, the value of field strength $E_{\text{ion}}$ begins to saturate, leading to the observed harmonic yield scaling of $\propto \tau^{-4.1}$ with pulse duration.

\begin{figure}[t]
\centering\includegraphics[width=0.95\columnwidth]{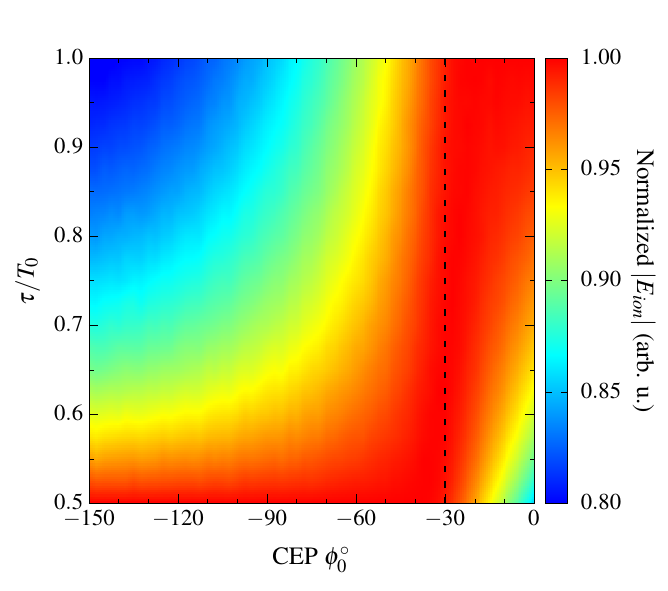}
\caption{Variation of the electric field strength $E_{\text{ion}}$ with laser pulse duration $\tau$ for different CEP $\phi_0$ values. The black dashed-line indicating nearly constant values of $E_{\text{ion}}$ with varying $\tau$ and fixed $\phi_0=-30^\circ$. The $E_{\text{ion}}$ values are normalized to their respective maximum for each CEP (i.e., normalized column-wise).}
\label{fig7}
\end{figure}

The abovementioned analysis suggests that the variation in harmonic yield with pulse duration is primarily influenced by the field strength $E_{\text{ion}}$ as the electron propagation time $T_{\text{prop}}$ remains nearly constant. Based on this observation, we calculated the variation of field strength $E_{\text{ion}}$ with pulse duration for different CEP $\phi_0$ values, as shown in Fig. \ref{fig7}. The $E_{\text{ion}}$ values are normalized to their respective maximum for each CEP (i.e., normalized column-wise). The trends in $E_{\text{ion}}$ for $\phi_0 = 0^\circ$ and $\phi_0 = -90^\circ$ (or $90^\circ$) exhibit opposite behavior, with one increasing and the other decreasing as pulse duration increases, in agreement with the harmonic yield results discussed earlier. Additionally, Fig. \ref{fig7} shows that for $\phi_0 \sim -30^\circ$, the field strength $E_{\text{ion}}$ remains unchanged as the pulse length increases, suggesting that the HHG yield would remain constant with varying pulse durations at $\phi_0 = -30^\circ$. It should be noted that the harmonic yield trends indicated in Fig. \ref{fig7} are applicable only when comparing different pulse durations for a fixed CEP value.

\begin{figure}[t]
\centering\includegraphics[width=1\columnwidth]{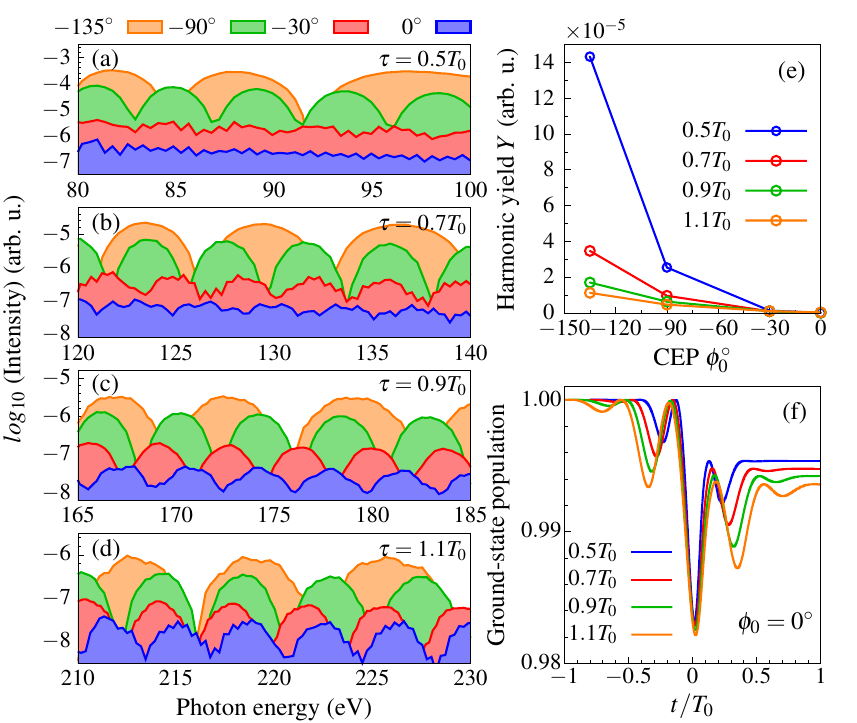}
\caption{HHG spectra generated for CEP $\phi_0$ values $0^\circ$ (blue), $-30^\circ$ (red), $-90^\circ$ (green), and $-135^\circ$ (orange) with fixed laser intensity $I_0 = 8\times 10^{14}$ W$/$cm$^2$ for different pulse durations $\tau$: (a) $0.5T_0$, (b) $0.7T_0$, (c) $0.9T_0$, and (d) $1.1T_0$. The harmonic yield in the harmonic energy range shown in panels (a-d) for each pulse duration is calculated as a function of CEP $\phi_0$ and shown in (e). (f) Temporal variation of the ground-state population of He atom for different pulse durations.}
\label{fig8}
\end{figure}

We also investigated the impact of CEP $\phi_0$ on the integrated harmonic yield $Y$ for various durations of sub-cycle driving fields. Figs. \ref{fig8}(a)-(d) show the HHG spectra (without any offset) generated for different CEP values and pulse durations. It is clear that the harmonic intensity increases as the CEP shifts from $0^\circ$ to $-135^\circ$ for all pulse duration cases. The harmonic yield for each pulse duration, calculated as a function of CEP $\phi_0$, is displayed in Fig. \ref{fig8}(e). A fixed harmonic bandwidth $\Delta \epsilon = 20$ eV [refer to Eq. \ref{yield}] is used for yield calculation, with the lower ($\epsilon_i$) and upper ($\epsilon_f$) energy limits corresponding to the photon energy ranges shown in Figs. \ref{fig8}(a)-(d). As indicated by the HHG spectra, the harmonic yield increases with CEP as it varies from $0^\circ$ to $-135^\circ$. Additionally, the rate of increase in harmonic yield becomes steeper as the pulse duration decreases, indicating that the impact of CEP variation is more pronounced for shorter pulse durations.

Furthermore, Fig. \ref{fig8}(f) shows the temporal variation of the ground-state population of the target atom for various pulse durations. Despite the laser pulses maintaining the same peak intensity, ground-state depletion is reduced for shorter pulses. This suggests that the target atom can endure higher laser intensities when driven by sub-cycle pulses, indicating that such pulses could enhance the harmonic yield, which would be beneficial for generating intense isolated attosecond pulses.

\begin{figure}[t]
\centering\includegraphics[width=1\columnwidth]{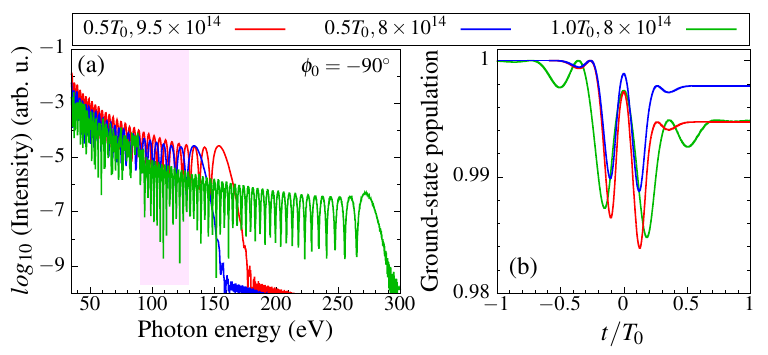}
\caption{(a) Comparison of HHG spectra generated for pulse durations $\tau=0.5T_0$, and $1.0T_0$, with CEP $\phi_0=-90^\circ$. (b) Temporal variation of the ground-state population of the target atom.}
\label{fig9}
\end{figure}

Finally, we address the critical question of whether sub-cycle pulses ($\tau<1.0T_0$) are more effective for harmonic generation compared to the single-cycle pulses ($\tau=1.0T_0$). To investigate this, we compare the results for pulse durations of $\tau=0.5T_0$ and $1.0T_0$. The analysis of harmonic yield in Sec. \ref{subsec2} suggests that a $0.5T_0$ pulse could provide a higher HHG yield, provided the CEP $\phi_0 < -30^\circ$. Fig. \ref{fig9}(a) compares the HHG spectra generated by $0.5T_0$ and $1.0T_0$ pulses. The harmonic yield is calculated within the energy range of $90-130$ eV (shaded region in Fig. \ref{fig9}(a)) for pulses of similar peak intensity $I_0 = 8 \times 10^{14}$ W/cm$^2$. The yield ratio for the $0.5T_0$ to $1.0T_0$ pulse case is $\sim11$, indicating that the $0.5T_0$ pulse produces harmonics with intensity more than an order of magnitude higher than the $1.0T_0$ pulse. Additionally, Fig. \ref{fig9}(b) shows the temporal variation of the ground-state population of a He atom for both pulse durations, where the $0.5T_0$ pulse causes less ground-state depletion than the $1.0T_0$ pulse, suggesting that the ground-state of the target atom can withstand higher laser intensity with a $0.5T_0$ pulse. We also calculated the HHG spectra for a $0.5T_0$ pulse with increased intensity ($I_0 = 9.5 \times 10^{14}$ W/cm$^2$) such that its ground-state depletion matches that of the $1.0T_0$ pulse. In this case, the yield ratio between the $0.5T_0$ pulse and the $1.0T_0$ pulse increases to approximately 18. 
Thus, $0.5T_0$ pulses can achieve atleast one order of magnitude higher HHG yield compared to $1.0T_0$ pulses when $\phi_0 < -30^\circ$.

Furthermore, the HHG spectra shown in Fig. \ref{fig2}(a) reveal that the harmonic continuum bandwidth for the $0.5T_0$ pulse is about $65\%$ of the total spectrum, whereas for the $1.0T_0$ pulse, it is approximately $48\%$. A broader harmonic continuum is crucial for generating short-duration isolated attosecond pulses (IAPs). However, the attochirp ($\beta$) for the $0.5T_0$ pulse is around 3.85 as/eV, higher than the 2.72 as/eV obtained for the $1.0T_0$ pulse, as shown in Fig. \ref{fig2}(c). This attochirp can be compensated by propagating the harmonics through dispersive media \cite{Mairesse2003_Science, Ko2012_JPhysB}. Additionally,
the harmonic cutoff energy is significantly reduced for sub-cycle fields due to the self-induced blueshift of the driving frequency. This reduction in cutoff energy can be partially mitigated by increasing the peak intensity of the sub-cycle pulse, as the ground-state depletion is less for shorter pulses. As evident from the Fig. \ref{fig9}(a), the harmonic cutoff is increased from $\sim 135$eV to $\sim 160$ eV, when the intensity of $0.5T_0$ pulse is increased such that the ground-state depletion is similar to the $1.0T_0$ duration pulse. Consequently, sub-cycle pulses offer a better opportunity for generating intense isolated attosecond pulses compared to the single-cycle driving fields.

\section{Summary and Concluding Remarks}
\label{sec4}

In summary, we investigated the effects of sub-cycle driving fields on high-order harmonic generation, focusing specifically on the influence of intrinsic chirp, carrier-envelope phase, and pulse duration. For the numerical modeling of sub-cycle pulse, we relied on the analytical expressions of sub-cycle pulsed beam, which are the exact solution of Maxwell's equations. We first analyzed the dependence of instantaneous laser frequency on pulse duration, revealing that intrinsic chirp induces a blueshift in the center frequency, scaling as $\propto \tau^{-5/4}$ with pulse length. The scaling is crucial in determining the harmonic cutoff energy and the emission properties of generated harmonics. A modified expression of the cutoff energy accounting for the intrinsic chirp, is presented in Eq. \ref{new_cutoff}. The attochirp $\beta$, a critical parameter for characterizing attosecond pulses, was found to scale similarly with pulse duration \(\beta \propto \tau^{-5/4}\). This scaling was verified for both CEP values of \(0^\circ\) and \(-90^\circ\), demonstrating its universality. Additionally, our analysis revealed that the harmonic cutoff energy and the duration of the synthesized attosecond pulses are both influenced by the intrinsic chirp of the sub-cycle pulses, with shorter pulses leading to reduced cutoff energies and longer attosecond pulse durations.

Moreover, we examined the scaling of harmonic yield with the number of cycle of the driving laser, which revealed CEP-specific trends. For \(\phi_0 = 0^\circ\), the yield increased as \(\tau^{5/4}\), whereas for \(\phi_0 = -90^\circ\), it decreased with a scaling of \(\tau^{-4.1}\). These findings highlight the critical role of both intrinsic chirp and CEP in the optimization of HHG processes driven by ultrashort pulses.

In conclusion, our findings indicate that sub-cycle pulses present considerable advantages over single-cycle pulses for producing IAPs via HHG. Sub-cycle pulses allow for a higher HHG yield because the target atom can endure greater laser intensities, thereby facilitating the creation of high-intensity IAPs. Additionally, they have a higher contribution of the harmonic continuum region in the total HHG spectra, which is essential for generating short-duration IAPs. Although sub-cycle pulses exhibit increased attochirp, this can be compensated by propagating the generated harmonics through dispersive materials \cite{Mairesse2003_Science, Ko2012_JPhysB}, making the sub-cycle pulses a viable option for generating intense IAPs of Fourier transform limited duration.

Our study provides valuable insights into high harmonic generation and attosecond pulse production using sub-cycle driving fields, offering a framework for controlling and understanding the temporal structure and efficiency of emitted harmonics. These results have significant implications for the development of ultrafast light sources and advancing the understanding of ultrafast processes on attosecond timescales.
 
\section*{Acknowledgments} Authors would like to acknowledge the financial support from the Ministry of Education, Culture, Sports, Science and Technology of Japan (MEXT) through Grants-in-Aid under grant no. 21H01850, and the MEXT Quantum Leap Flagship Program (Q- LEAP) (grant no. JP-MXS0118068681). This project was supported by the RIKEN TRIP initiative (Leading-edge semiconductor technology).
  
\bibliographystyle{apsrev4-2}
\bibliography{Bibliography}

\end{document}